\documentclass[pdflatex,sn-mathphys-num]{sn-jnl}

\usepackage{amsmath,amssymb,amsfonts}%
\usepackage{booktabs}%

\begin{document}
	
	\title[Article Title]{
		Telecom C-band single-photon sources with a semiconductor-dielectric microresonator
	}
	
	
	\author*[1]{\fnm{Yuriy} \sur{Serov}}\email{serovjurij@beam.ioffe.ru}
	
	\author[1]{\fnm{Aidar} \sur{Galimov}}\email{galimov@mail.ioffe.ru}
	
	\author[1]{\fnm{Sergey} \sur{Sorokin}}\email{sorokin@beam.ioffe.ru}
	
	\author[1]{\fnm{Nikolai} \sur{Maleev}}\email{maleev@beam.ioffe.ru}
	
	\author[1]{\fnm{Marina} \sur{Kulagina}}\email{marina.kulagina@mail.ioffe.ru}
	
	\author[1]{\fnm{Yuriy} \sur{Zadiranov}}\email{zadiranov@mail.ioffe.ru}
	
	\author[1]{\fnm{Grigorii} \sur{Klimko}}\email{klimko@beam.ioffe.ru}
	
	\author[1]{\fnm{Maxim} \sur{Rakhlin}}\email{rakhlin.maxim@mail.ioffe.ru}
	
	\author[1]{\fnm{Alexey} \sur{Veretennikov}}\email{veretennikov.a@mail.ioffe.ru}
	
	\author[1]{\fnm{Gleb} \sur{Veyshtort}}\email{gleb\_veysh@mail.ru}
	
	\author[1]{\fnm{Olga} \sur{Lakuntsova}}\email{o.e.lakuntsova@mail.ioffe.ru}
	
	\author[1]{\fnm{Yuliya} \sur{Salii}}\email{guseva.julia@mail.ioffe.ru}
	
	\author[1]{\fnm{Daria} \sur{Berezina}}\email{dariya.burenina@mail.ioffe.ru}
	
	\author[1]{\fnm{Sergey} \sur{Troshkov}}\email{s.troshkov@mail.ioffe.ru}
	
	\author[1]{\fnm{Demid} \sur{Kirilenko}}\email{demid.kirilenko@mail.ioffe.ru}
	
	\author[1]{\fnm{Alexey} \sur{Blokhin}}\email{aleksey.blokhin@mail.ioffe.ru}
	
	\author[1]{\fnm{Alexei} \sur{Vasil’ev}}\email{vasiljev@mail.ioffe.ru}
	
	\author[1]{\fnm{Alexander} \sur{Kuzmenkov}}\email{kuzmenkov@mail.ioffe.ru}
	
	\author[1]{\fnm{Mikhail} \sur{Bobrov}}\email{bma@mail.ioffe.ru}
	
	\author[1]{\fnm{Irina} \sur{Sedova}}\email{irina@beam.ioffe.ru}
	
	\author[1]{\fnm{Tatiana V.} \sur{Shubina}}\email{shubina@beam.ioffe.ru}
	
	\author[1]{\fnm{Alexey A.} \sur{Toropov}}\email{toropov@beam.ioffe.ru}

	
	
	\affil[1]{\orgname{Ioffe Institute}, \orgaddress{\street{26 Polytechnicheskaya St.}, \city{Saint Petersburg}, \postcode{194021}, \country{Russia}}}
	
	
	\abstract{
		Secure communications with quantum key distribution over fiber-optic links is one of the few recognized applications of quantum physics at the level of individual quanta---single C-band photons. Currently, the widely used sources of such photons are highly attenuated laser pulses, featured by a low probability of single
		photon occurrence. Here, we present an efficient source with an $\text{InAs/GaAs}$ quantum dot on a metamorphic buffer layer inside a micropillar-shaped microcavity. The key innovation is the use of different semiconductor and dielectric materials to form the lower (GaAs/AlGaAs) and upper (Si/SiO$_2$) Bragg reflectors. Compatibility of these materials in a monolithic source is achieved by depositing a small amount of Si/SiO$_2$ pairs on an incomplete micropillar made from a coherent heterostructure grown by molecular beam epitaxy. This design enables resonant excitation with $\pi$-pulses and generation of polarized photons with a record-breaking end-to-end efficiency of 11\%.
	}
	

	
	
	
	\maketitle
	
	To ensure secure telecommunication over fiber-optic channels using the concept of quantum	key distribution \cite{Bennett1984, Xu2020}, efficient single-photon sources (SPSs) for the telecommunication C-band	(1530–1565 nm) are required, the experimental implementation of which is a complex and not yet fully resolved problem \cite{Holewa2025}. Currently, sources of highly attenuated laser pulses are used, despite the fundamental limitation of $1/e$ ($\approx0.37$) on the probability of producing a single photon in a Poisson pulse \cite{Vajner2022}. This is driving interest in C-band SPSs based on semiconductor quantum dots (QDs) inside a microresonator \cite{Takemoto2015, Vajner2022, Holewa2025}, which have the potential to provide much higher brightness. The most efficient SPSs currently developed for operation in the 900–1000~nm range are based on a micropillar structure with an InAs/GaAs QD and two distributed Bragg reflectors (DBRs). Such a microresonator can either be completely monolithic, combining two semiconductor DBRs within a columnar structure \cite{Ding2016, Wang2019}, or use one monolithic DBR and one external mirror as part of a tunable microcavity \cite{Tomm2021, Ding2025}. End-to-end efficiencies, defined as the probability of receiving one polarized photon in a single-mode fiber per pump pulse, for the best devices of this type exceed $0.7$ \cite{Ding2025, Zhang2025}. 
	
	However, the most efficient C-band photon source fabricated to date, using InAs/InGaAs QDs on an InGaAs metamorphic buffer layer (MBL) in a bull’s-eye resonator, has an efficiency of polarized single-photon emission to the fiber of only $0.064$ \cite{Nawrath2023, Yang2024}. That is insufficient to improve the performance of quantum key distribution schemes compared to attenuated laser pulses with decoy states \cite{Hwang2003, Zhao2006, Scarani2009}. As an alternative solution, it was proposed to use difference-frequency parametric conversion of 940 nm to the C-band in a periodically poled lithium niobate waveguide \cite{Morrison2021}, which provides an efficiency of about 0.082 \cite{Zahidy2024}.
	
	Notably, the epitaxial growth of heterostructures with QDs emitting in the C-band is currently quite well developed. Such structures can be successfully grown using the Stranski-Krastanov mode with a lattice mismatch between the QDs and the surrounding matrix of approximately 4\%, compared to the 7\% mismatch required to grow QDs emitting in the 900-1000 nm range. These can be either InAs/InP QD structures grown on InP substrates \cite{Miyazawa2016, Holewa2022} or InAs/InGaAs QDs grown on an InGaAs MBL on a GaAs substrate \cite{Sittig2022, Wyborski2023, Sorokin2024}. The main obstacle to creating efficient SPSs based on them is the impossibility of reproducing the resonator designs that are well developed for shorter wavelengths, in particular, columnar (pillar) microcavities with DBRs \cite{Ding2016, Wang2019}. Indeed, forming DBRs on InP substrates is challenging task due to the lack of lattice-matched materials with sufficiently high refractive index contrast. On GaAs substrates, it is possible to create a lattice-matched GaAs/AlGaAs bottom DBR structure, followed by an MBL structure with QDs inserted into a short cavity. However, the creation of a second, upper	monolithic semiconductor structure of Al$_{0.9}$Ga$_{0.1}$As/GaAs DBR remains an unsolved problem due	to the mismatch of its crystal lattice with the upper InGaAs layer and the impossibility of inserting another correcting MBL. This dictates difficulties in producing C-band SPS with high brightness and high single-photon indistinguishability \cite{Nawrath2023, Nawrath2021, Joos2024, Holewa2024, Vajner2024, Kim2025, Hauser2026}.
		
	In this paper, we present a new, highly efficient micropillar-shaped source of single C-band photons, that utilizes different material systems for the lower and upper DBRs in the micropillar. The lower DBR is a multilayer GaAs/AlGaAs semiconductor structure, while the upper DBR consists of several pairs of quarter-wavelength Si/SiO$_2$ layers. We demonstrate the compatibility of these two material systems in a single monolithic microresonator and the applicability of molecular beam epitaxy (MBE) for growing InAs/InGaAs QDs on a high-quality MBL that is sufficiently thin to fit within a three-wavelength-thick resonator cavity. The manufactured SPS with this design enabled the efficient coherent resonant $\pi$-pulse pumping mode---previously unused for QD/cavity systems, emitting in this spectral range. It demonstrated an end-to-end efficiency of $0.11$, the best value reported to date for both monolithic C-band photon sources and those based on nonlinear frequency conversion. The generated single-photon radiation has a purity of $0.96$ and exhibits a photon indistinguishability of up to 38\%, which is a fairly high value for high-efficiency QD-based C-band sources.
	
	\section{SPS design and characterization}
	
\begin{figure}[t]
		\centering
		\includegraphics[width=0.97\textwidth]{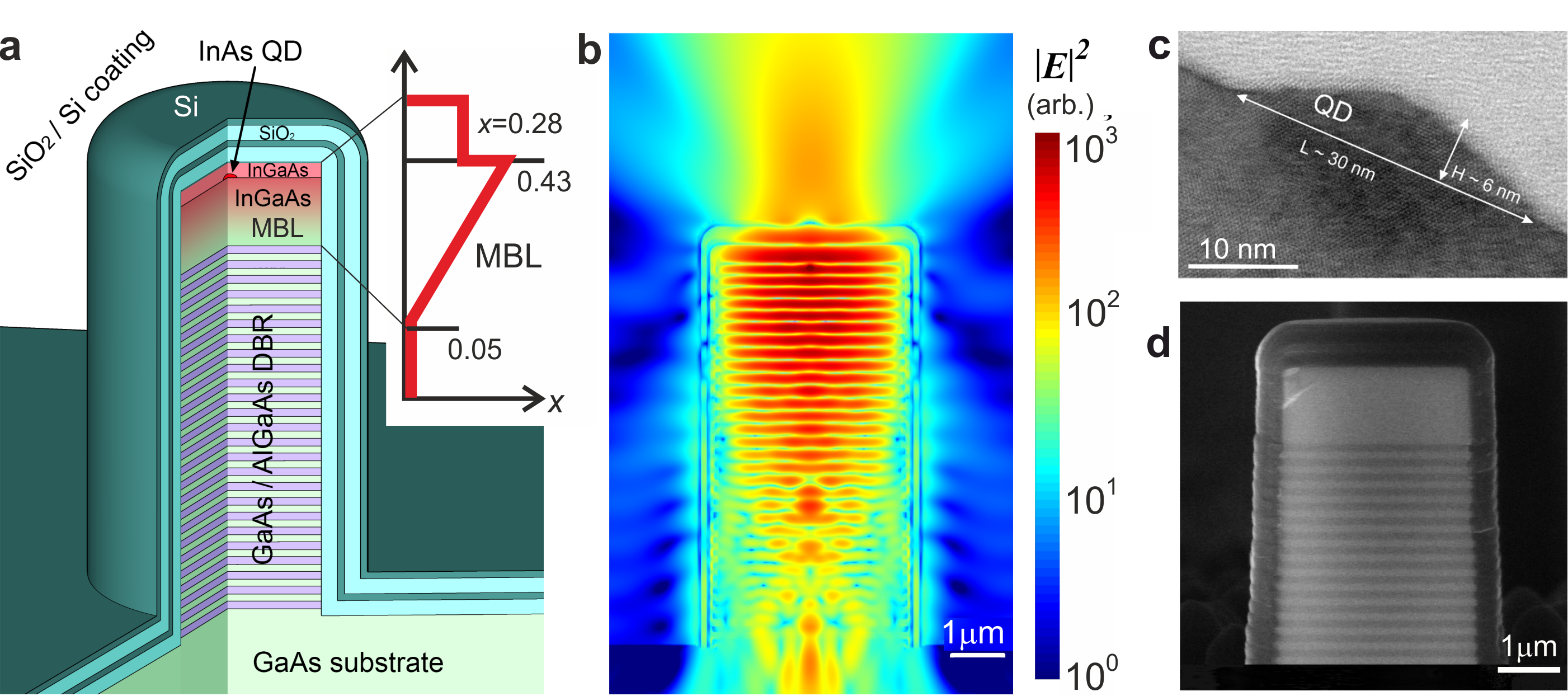}
		\caption{$\vert$ \textbf{Conceptual design of SPSs with a semiconductor-dielectric microresonator.} \textbf{a}, Schematic of the micropillar structure with two DBRs made of AlGaAs/GaAs (lower) and Si/SiO$_2$ (upper). The Si/SiO$_2$ coating covers both the top and side walls. The inset shows a linear gradient of the In content $x$ in the In$_x$Ga$_{1-x}$As MBL varying from $0.05$ to $0.43$ to decrease the lattice mismatch to 4\%. \textbf{b}, Optical field energy distribution in a 4-$\mu$m diameter micropillar structure obtained by FDTD simulations for 25 pairs of $\lambda/4$ GaAs/AlGaAs layers in the lower DBR and 2 pairs of $\lambda/4$ Si/SiO$_2$ layers in the upper and sidewall DBR. \textbf{c}, Transmission electron microscopy (TEM) image of a typical QD emitting in the C-band. \textbf{d}, Scanning electron microscopy (SEM) image of cleaved cross-section of a micropillar with the composite microresonator.
		}\label{fig1}
	\end{figure}

	The concept of the proposed semiconductor-dielectric micropillar resonator is illustrated in Fig.~\ref{fig1}a. With a typical number of GaAs/AlGaAs layer pairs in the lower DBR of about 25 \cite{Rakhlin2023}, only two Si/SiO$_2$ pairs are required in the upper DBR to provide a sufficiently high reflectivity (above $0.95$) throughout the whole C-band. The Si/SiO$_2$ layers of the DBR cover both the top and side walls of the micropillar. Finite-difference time-domain (FDTD) simulations show that the sidewall DBR suppresses lateral light leakage but increases the optical mode volume that slightly reduces the Purcell enhancement. Both the MBL and QDs are placed inside the $3\lambda$-cavity between the two DBRs. Calculated optical field energy distribution exhibits a narrow beam pattern typical for a perfect micropillar resonator (Fig.~\ref{fig1}b).
	
	The parent heterostructure for the SPS was grown by MBE on a GaAs substrate. The inset in Fig.~\ref{fig1}a shows the MBL profile providing optimal conditions for the formation of self-organized InAs QDs emitting near 1.55~$\mu$m, and a typical QD shape is illustrated in Fig.~\ref{fig1}c. Periodically arranged micropillars with various diameters were fabricated using photolithography and reactive ion-plasma etching. They were then coated with a Si/SiO$_2$ DBR using ion-assisted reactive magnetron sputtering. Figure~\ref{fig1}d shows an image of a cleaved micropillar revealing clearly resolved multilayer DBR structures. Details on the sample fabrication are given in Methods.
	
\begin{figure}[t]
		\centering
		\includegraphics[width=0.97\textwidth]{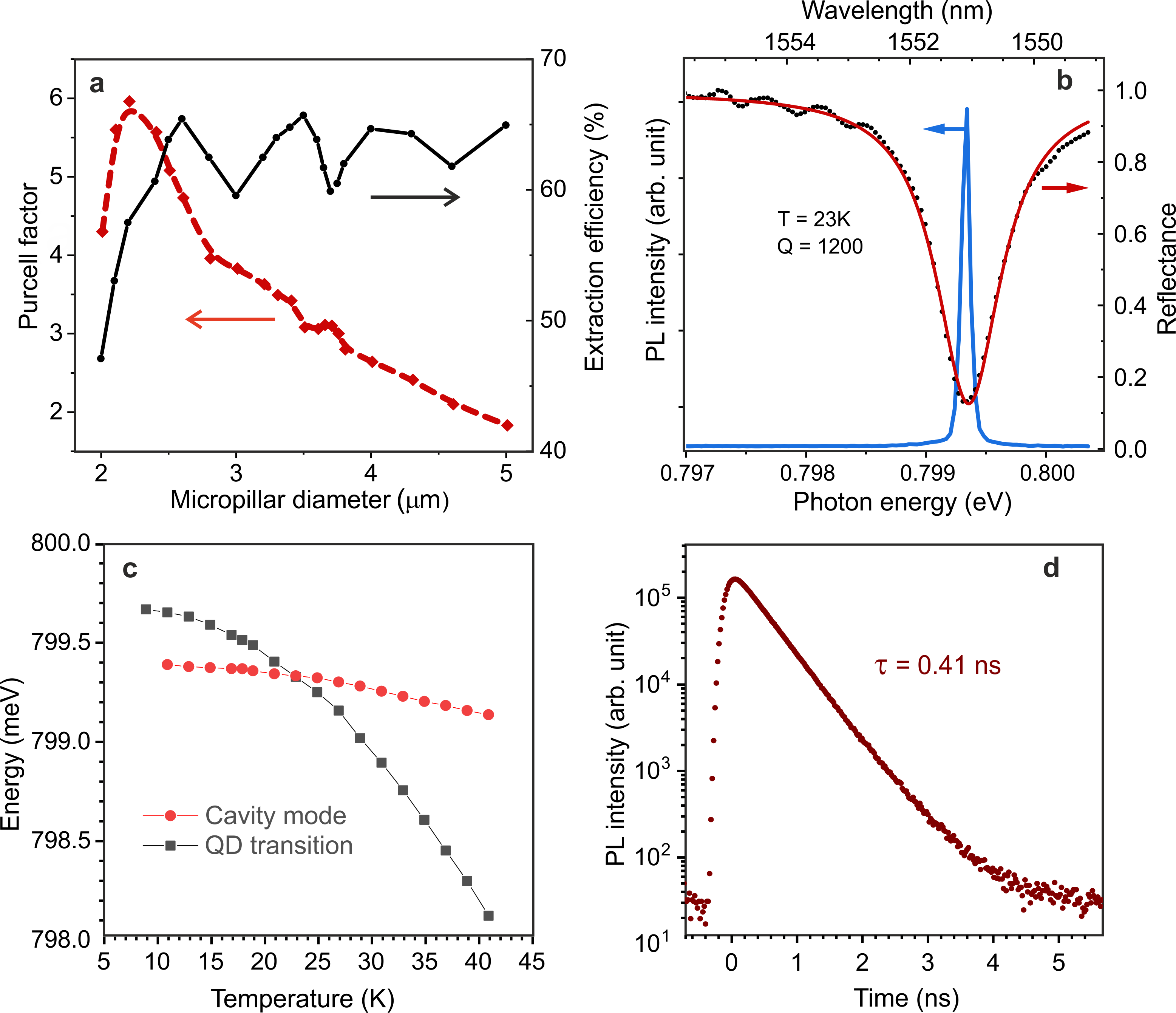}
		\caption{$\vert$ \textbf{Basic optical properties of SPS with a composite microresonator.} \textbf{a}, Calculated Purcell factor and photon extraction efficiency at numerical aperture $NA = 0.68$ (right axis) as functions of the micropillar diameter. \textbf{b}, Typical reflectance and resonant PL spectra of the studied samples. \textbf{c}, Temperature dependencies of the cavity mode and PL line energies for a sample with an operating temperature of 23~K. \textbf{d}, Typical decay curves of the resonant PL.
		}\label{fig2}
	\end{figure}

To test the SPS characteristics, optical measurements were performed on micropillars with diameters $d$ in the range of $3.5$--$4$~$\mu$m. This choice represents a trade-off: the Purcell factor $F_p$ decreases with the increasing diameter, but the photon extraction efficiency $\eta_\text{out}$---a key figure of merit---drops for small $d$ (Fig.~\ref{fig2}a). Furthermore, the presence of oscillations in the $\eta_\text{out}(d)$ dependence, revealed by FDTD simulation, implies the need for an even more precise selection of the micropillar diameter. In particular, the maximum in the chosen diameter range in Fig.~\ref{fig2}a corresponds to an extraction efficiency $\eta_\text{out}>65\%$, a Purcell factor $F_p \approx 3$, and a cavity quality factor $Q \approx 1200$. More precise modeling, which takes into account the presence of structural defects in real structures, indicates that such defects affect primarily the extraction efficiency and could decrease it to 45\%. Details of the FDTD simulations are provided in the Methods section.

	Typical reflectivity and photoluminescence (PL) spectra of the studied SPSs are shown in Fig.~\ref{fig2}b. The dip in reflectivity corresponding to the fundamental optical mode of the microresonator is located at a wavelength of 1551 nm. Its full width at half maximum (FWHM) is $720$~$\mu$eV, which is significantly wider than the Fourier-limited PL linewidth ($<100$~$\mu$eV). The $Q$-factors of various microresonator samples range from 1100 to 1200 in agreement with the FDTD simulation. To precisely tune the emission line energy to the resonator mode, we used the difference in the temperature dependencies of their positions (Fig.~\ref{fig2}c). The intersection of these dependencies determines the operating temperature of a particular SPS.
	
	The PL decay curve in Fig.~\ref{fig2}d was measured under resonant excitation of the SPS at the operating temperature, which ensures a minimal decay time. The absence of quantum beat oscillations in the decay curve allows us to attribute the PL line to the recombination of charged exciton (trion) in a QD \cite{Ollivier2020}. We estimated the emission lifetime to be $0.46\pm0.05$~ns in complete microcavities. With the FDTD Purcell factor $F_p \approx 3$, the lifetime of InAs QDs on MBL without such a microcavity is about $1.4$~ns in good agreement with the typical values previously reported \cite{Paul2017, Nawrath2023}.

	\section{C-band single-photon emission}
	
	Single-photon emission characteristics were measured under resonant excitation by $\pi$-pulses of an SPS located in a helium-flow cryostat. The single-photon statistics was determined by measuring the second-order autocorrelation function $g^{(2)}(t)$ in a Hanbury Brown-Twiss (HBT) setup. To characterize photon indistinguishability, we measured the in-fiber Hong-Ou-Mandel (HOM) interference visibility $V_\text{HOM}$. A detailed description of the setup and measurements is provided in the Methods section.
	
	\subsection{Brightness}
	
\begin{figure}[h]
		\centering
		\includegraphics[width=0.97\textwidth]{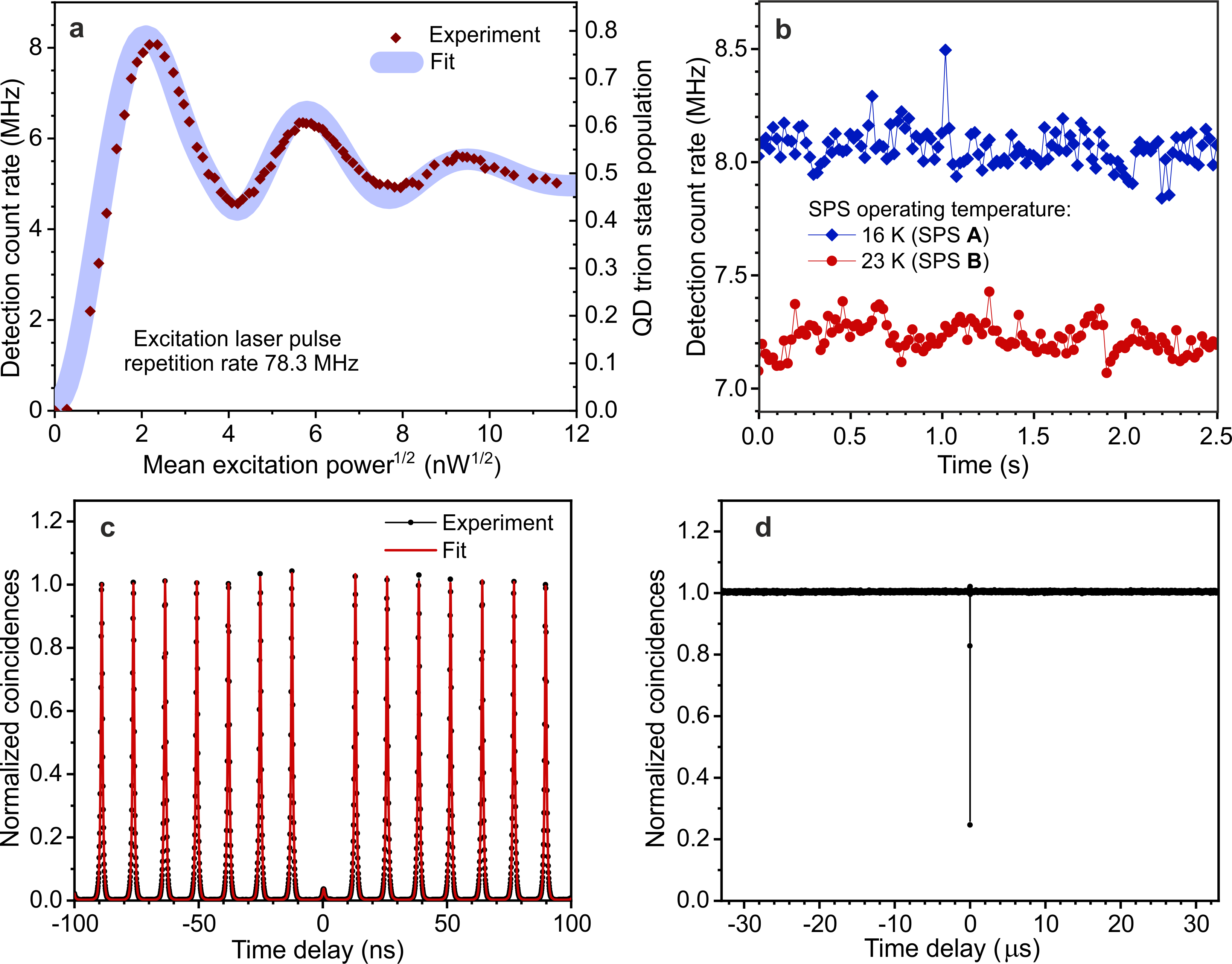}
		\caption{$\vert$ \textbf{Single-photon emission from an SPS: Brightness and Purity.}
			\textbf{a}, The single-photon detection rate under resonant pulse excitation (Rabi oscillations) with a fit. \textbf{b}, The single-photon detection rate over time for two sources with different operating temperature excited by $\pi$-pulses. \textbf{c} and \textbf{d}, Autocorrelation function $g^{(2)}(t)$ of the SPS emission measured under $\pi$-pulse excitation on different time scales: short (panel \textbf{c}) and long (panel \textbf{d}). The peaks are not visible in \textbf{d} due to the large time bin of the histogram, which is equal to the repetition period of the laser pulses. No blinking is observed over long intervals.}
		\label{fig3}
	\end{figure}

	Under resonant excitation, the recorded photon count rate exhibits typical Rabi oscillations with a pulse area proportional to the square root of the average excitation power (Fig.~\ref{fig3}a). This dependence reflects the coherent dynamics of the QD excited state population, which reaches a maximum at a $\pi$-pulse. The observed damping and slight aperiodicity of the oscillations can be attributed to exciton-phonon interaction \cite{Forstner2003, Wigger2018, Ramsay2010}. The oscillation intensity asymptotically decreases with increasing power to a value corresponding to an excited state population probability of $0.5$ \cite{Forstner2003, Monniello2013}. Using this value, we determined the excited state preparation fidelity for a $\pi$-pulse to be $\eta_\text{exc} = (77 \pm 4)\%$.
	
	Figure~\ref{fig3}b shows the photon count rate as a function of time, measured under resonant excitation with a $\pi$-pulse for two different samples, labeled  \textbf{A} and \textbf{B}, to illustrate the dispersion of this parameter. Although their operating temperatures significantly differ, the average difference in maximum emission intensities is only about 10\% ($8.065\pm0.022$~MHz versus $7.226\pm0.018$~MHz), and both SPSs \textbf{A} and \textbf{B} demonstrate reasonably high single-photon purity of $g^{(2)}(0)=0.043$ and $0.059$, respectively, under $\pi$-pulse excitation. Sample \textbf{A} shows the characteristics closest to the idealized FDTD calculations and best demonstrates our concept. The results of detailed studies of this sample are presented below.
	
	We define the unpolarized first-lens brightness $\eta_\text{FL}$ as the mean probability of detecting a single photon of arbitrary polarization at the first lens per excitation pulse. In a symmetric microcavity, the trion state emits photons with equal probability for both linear polarizations. Therefore, the brightness of arbitrarily polarized emission at the first lens is twofold higher than that of the specifically polarized emission $\eta_\text{FL\,pol}$. The end-to-end efficiency of single-photon generation $\eta_\text{end}$ is defined here as the mean probability of obtaining a polarized single photon in the optical fiber per excitation pulse, since the measurements were performed under cross-polarization filtering conditions. These SPS parameters were calculated from the detection rate in Fig.~\ref{fig3}b, taking into account the efficiency of the superconducting nanowire single-photon detector (SNSPD), dark counts, and the optical setup transmittance $\eta_\text{opt}=(50\pm2)\%$ (see Methods for details). We derived the maximum values of $\eta_\text{end}=(11.0\pm0.3)\%$ and $\eta_\text{FL} = (44\pm3)\%$ under $\pi$-pulse excitation.
	
	The corresponding light extraction efficiency $\eta_\text{out}$ was also estimated from the expression $\eta_\text{FL} = \eta_\text{exc} \cdot \eta_\text{blink} \cdot \eta_\text{out}$ \cite{Somaschi2016}. Here $\eta_\text{blink}$ is the stability of QD bright state associated with the blinking process, which is as high as $\approx95\%$, as will be discussed below, and $\eta_\text{exc} = (77 \pm 4)\%$ for the same structure was estimated from Fig.~\ref{fig3}a. The obtained value $\eta_\text{out} = (60\pm8)\%$ agrees well with the calculated values and indicates small structural defects density in the sample. In both this estimate and the FDTD simulations, we include the QD-to-mode coupling efficiency $\beta=F_p/(F_p + 1)$ in the value of $\eta_\text{out}$.

	\subsection{Single-photon purity}
	
	Figure~\ref{fig3}c shows the typical $g^{(2)}(t)$ function, measured for the SPS \textbf{A} under $\pi$-pulse excitation. The side peaks of the histogram were normalized to unity. The dependence demonstrates pronounced antibunching at the zero peak. The $g^{(2)}(0)=0.043\pm 0.002$ value was defined as the zero peak area normalized to the average area of the side peaks, evaluated from the fitting, as described in the Methods section. With decreasing excitation power, this parameter can be improved, but at the expense of SPS brightness. For example, a 30\% decrease in count rate yields $g^{(2)}(0) = 0.030\pm0.002$, indicating flexibility in the SPS performance.
	
	Near the main dip, weak bunching is observed, usually attributed to blinking of the QD charge state \cite{Hilaire2020}. For this SPS, such blinking is very small and occurs over an atypically short time scale of about 20~ns. A longer-term correlation measurement (Fig.~\ref{fig3}d) shows that the QD exhibits no other blinking mechanisms. From the data fit presented in Fig.~\ref{fig3}c, we conclude that the QD in the SPS remains in the bright state for $\eta_\text{blink}=95\%$ of the time.

	\subsection{Single-photon indistinguishability}
	
\begin{figure}[b]
		\centering
		\includegraphics[width=0.97\textwidth]{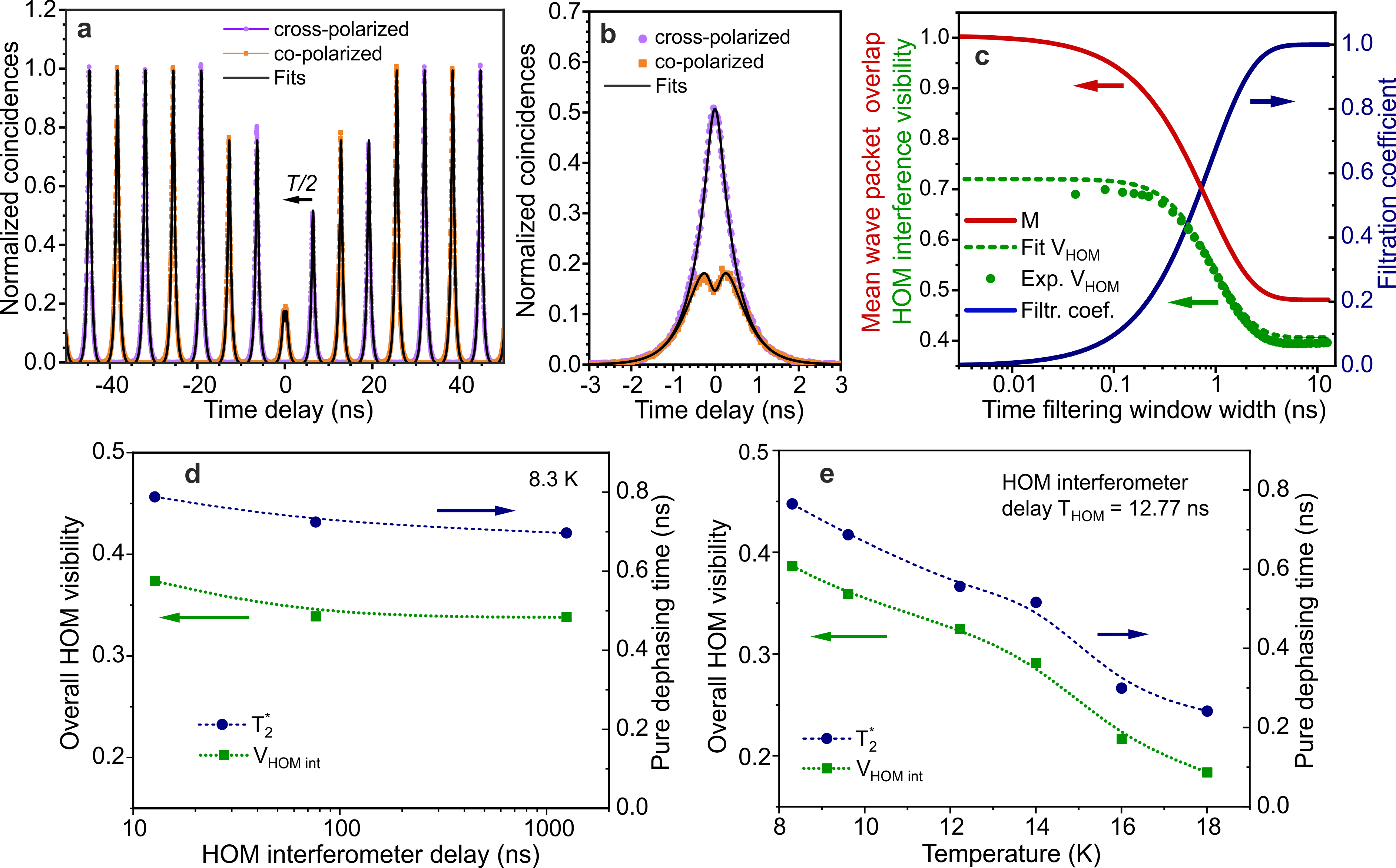}
		\caption{\textbf{Single-photon indistinguishability measurements.}
			\textbf{a}, HOM histograms for cross- and co-polarized photon counts at 8.3 K with a delay $T_\text{HOM}$ in the interferometer equal to the laser pulse repetition period $T$. The cross-polarized signal is shifted along the time axis by $T/2$. \textbf{b}, Detailed view of the central peaks of the histograms in \textbf{a}. \textbf{c}, Time-filtered HOM interference visibility $V_\text{HOM}$, the mean photon wave packet overlap $M$, and the filtering coefficient as a function of the filtering window width. The green dashed line corresponds to the values calculated by integrating the fitting lines in \textbf{b}; the green dots are obtained by directly summing the experimental data in the filtering window after background subtraction. The filtering coefficient indicates the fraction of the signal within the filtering window. \textbf{d}, Dependencies of the fitting parameter $T^*_2$ and the integrated interference visibility $V_\text{HOM}$ on the interphoton delay $T_\text{HOM}$. \textbf{e}, $T^*_2$ and $V_\text{HOM}$ as functions of temperature. Lines in \textbf{d} and \textbf{e} are shown only to guide the eye.}
		\label{fig4} 
	\end{figure}

	HOM characterization measurements were performed using coherent $\pi$-pulse excitation with additional investigation of temperature variation effects. The highest indistinguishability was achieved at the decreased temperature of $8.3$~K, but at the cost of reduced brightness due to deviation from the optimal operating regime. The corresponding HOM histograms are shown in Fig.~\ref{fig4}a, where the trace for orthogonally polarized photons is time-shifted by $T/2$ ($T$ is the pulse excitation period). A magnified view of the central peaks is shown in Fig.~\ref{fig4}b, where a clear difference in the peak heights between co- and cross-polarized measurements stems from single-photon interference, which occurs when the photon polarizations are the same in the interferometer arms. The peak areas were estimated by integrating the signal over a 4~ns time interval around zero delay after background subtraction. The ratio of these peak areas yields an integrated HOM interference visibility of $V_\text{HOM}=38$~\%.
	
	The pronounced dip in the central peak in Fig.~\ref{fig4}b clearly indicates rapid photon dephasing along with a high degree of indistinguishability for time-filtered photons. In an idealized case, the shape of the central peak can be described by the equation \cite{Holewa2024, Kim2025, Bylander2003}:
	\begin{equation}
		I(t)=\left[A \cdot \exp \left(-\frac{|t|}{T_1}\right)   \left(1-M_0 \cdot \exp\left(-\frac{2\cdot |t|}{T^*_2}\right) \right) \right] \ast I_\text{instr}(t),
	\end{equation}
	where $T_1 = \tau$ is the trion lifetime, $T^*_2$ is the pure dephasing time, $M_0$ is the mean photon wave packets overlap exactly at zero detection delay, and the symbol $\ast$ denotes the convolution with the instrument response function $I_\text{instr}(t)$. This equation emphasizes the importance of decreasing the lifetime $T_1$ along with increasing the dephasing time $T^*_2$ to achieve high indistinguishability. We use this equation to fit the data including additional corrections for the nonzero value of $g^{(2)}(0)$, imperfectly balanced beam splitters in the HOM setup, and the non-unity classical interference visibility in the interferometer. In particular, from the curve fitting in Fig.~\ref{fig4}b, the values $T^*_2 = 0.79$~ns and $M_0 = 1$ was determined, as well as the corresponding coherence time $T_2 = 0.40$~ns, which is among the best values reported in this spectral range \cite{Holewa2025} (see  the Methods section for details).
	
	Figure~\ref{fig4}c shows the dependence of the time-filtered HOM interference visibility $V_\text{HOM}$ on the width of the temporal filtering window used in the experiment. The values of $V_\text{HOM}$ are limited by the interferometer imperfections and, for small filtering windows, by the width of the instrument response function. To characterize the SPS properties specifically, the mean photon wave packet overlap $M$, averaged over the filtering window, was derived from the fitting. For a small window width, it converges to the $M_0$ value and corresponds to almost perfect indistinguishability, but at the cost of a significant loss in brightness due to the filtered-out photons. Thus, the moderate overall HOM visibility, which is atypical for precise resonant excitation of a QD, is clearly associated with the short dephasing time. This result suggests that the main challenges in achieving high indistinguishability in the C-band are not related to excitation-induced temporal jitter. Rather, they could be related to spectral diffusion of the PL line or phonon-induced dephasing. The beneficial effects of improved structure quality \cite{Kim2025, Hauser2026} and additional spectral filtering \cite{Joos2024} point to a possible contribution from spectral diffusion. This may also contribute to the fast blinking observed in our $g^{(2)}(t)$ measurements.
	
	To discriminate between spectral diffusion and phonon-induced dephasing mechanisms, we followed the approach suggested in \cite{Thoma2016} and performed HOM measurements for different photon delays in the HOM setup and at different temperatures. The obtained dependencies of the overall interference visibility $V_\text{HOM}$ and the pure dephasing time $T^*_2$ on the delay time between photons and on temperature are shown in Figs. \ref{fig4}d and \ref{fig4}e, respectively, while $M_0$ remains close to unity within the error margins for all measurements. Clearly, $V_\text{HOM}$ and $T^*_2$ are virtually insensitive to photon delay, at least up to 1 $\mu$s, but rapidly degrade with increasing temperature. In particular, at the operating temperature of the structure, which ensures maximum brightness, the indistinguishability degrades from 38\% to 22\%, despite the shortening of the lifetime $T_1$ due to diminished detuning between the PL line and the cavity optical mode. This behavior indicates phonon-induced dephasing as the main reason for photon partial distinguishability in our structures, which could be overcome in the future by using SPSs with a lower operating temperature or a shorter excited state lifetime $\tau$.

	\section{Comparison of SPS parameters}
	
	Table~\ref{tab:comparison} compares the key parameters of our SPS with other sources that, prior to our work, had demonstrated at least one record-breaking parameter or an outstanding combination of parameters. These SPSs, which use a circular Bragg grating (CBG) as a resonator, are clearly inferior in brightness to our technologically simpler design. This new design achieves an efficiency of 11\% in generating polarized single photons at the end of a single-mode optical fiber, which is almost twice the previous record. In our measurement setup, this result corresponds to a nonpolarized "first lens" emission efficiency of 44\%.
	
	\begin{table}[h]
		\caption{Comparison of the single-photon emission properties of telecom C-band sources based on InAs/In(Al)GaAs QDs with record brightness or indistinguishability. SDMR - semiconductor-dielectric
			micropillar resonator; CBG~--- circular Bragg grating; LA~--- longitudinal acoustic phonon assisted excitation; SUPER~---swing-up of quantum emitter population; $\tau$~--- QD lifetime; $\eta_\text{end}$~--- total and polarized SPS end-to-end efficiencies.}
		\label{tab:comparison}
		\begin{tabular}{@{}llllllll@{}}
			\toprule
			Structure design & Resonator & Excitation & $\tau$, ps & $\eta_\text{end}$, $\%$\footnotemark[1] & $g^{(2)}(0)$ & $V_{\text{HOM}}, \%$ & Ref. \\
			\midrule
			MBL/GaAs & SDMR & resonant & 410  & --/11.0   & 0.043 & 38/22\footnotemark[2]  & this work \\
			MBL/GaAs & CBG & p-shell & 520  &  6.3/6.4\footnotemark[3]  & 0.007 & 8.1 & \cite{Nawrath2023, Yang2024} \\
			MBL/GaAs & CBG & LA, SUPER & 460  & 6.09/--    & 0.069 & 34.9  & \cite{Joos2024} \\
			InAlGaAs/InP & CBG & LA & 130  & 0.5/-- & 0.017 & 91.7  & \cite{Hauser2026} \\
			\bottomrule
		\end{tabular}
		\footnotetext[1]{Total brightness / brightness of polarized emission. The "--" sign means that the quantity was either not specified or could not be measured, as was the case for the cross-polarization scheme. In all cases, the brightness at the end of the optical fiber is given.}
		\footnotetext[2]{Data are presented for 8.3 K/16 K (operating temperature).}
		\footnotetext[3]{The first value corresponds to total brightness in \cite{Nawrath2023}, while the second---for polarized emission---was obtained in a subsequent work with an improved optical setup \cite{Yang2024}.}
	\end{table}
	
	The key factors enabling such brightness are resonant coherent excitation with $\pi$-pulses, which ensures high excited state preparation fidelity, and high photon extraction efficiency of the semiconductor-dielectric micropillar resonator compared to previously reported values for various cavity designs \cite{Holewa2025}. It should be noted that we observed a significant decrease in efficiency with the commonly used quasi-resonant excitation. We emphasize that this record brightness applies to the generation of polarized single photons, when the cross-polarization scheme filters out half of the total radiation. Even with the loss of 50\% of the signal, our design demonstrates single polarized photon collection efficiency of 22\% at the first lens, almost on par with the record value of 24\% for the elliptical CBG \cite{Ge2024}, where the filtered-out fraction of orthogonally polarized photons is below 2\%. Using a microresonator with an elliptical cross-section could further increase the efficiency \cite{Wang2019}.
	
	The obtained degree of indistinguishability is still far from unity; however, it is comparable to the best values for SPSs suitable for use in the telecom C-band. Higher values were reported only for sources with unspecified \cite{Kim2025} or comparatively low \cite{Hauser2026} emission efficiency. We demonstrated that in our SPS, the indistinguishability can be nearly doubled (38\% versus 22\%) by lowering the operating temperature, which suppresses phonon-induced dephasing. In general, achieving high indistinguishability primarily requires a higher Purcell factor to achieve a shorter PL lifetime of the QDs. Further acceleration of spontaneous emission can be achieved by shortening the MBL to fit within a 1$\lambda$ or 2$\lambda$ cavity \cite{Lakuntsova2026}. This will lead to a decrease in the volume of the confined optical mode and, consequently, an increase in the Purcell factor.
	
	\section{Conclusions}
	
	In summary, we have developed a new design and fabrication technology for a single-photon source intended for C-band telecommunication networks. The source is a monolithic semiconductor-dielectric micropillar resonator with a $3\lambda$ cavity containing a metamorphic InGaAs buffer layer with InAs/InGaAs QDs. Due to the sufficiently high Q-factor of the resonator and the low contribution of scattered light under cross-polarization filtering, this SPS enabled strictly resonant pumping with $\pi$-pulses, resulting in a record-breaking end-to-end efficiency for the C-band in generating polarized single photons with a relatively high degree of photon indistinguishability.
	
	The production of SPSs with such a composite resonator is technologically simpler than that of most other types. This design opens a direct path to further enhancing the SPS efficiency by using an anisotropic microcavity with a polarization-sensitive Purcell effect. Furthermore, the developed technology for epitaxial growth of parent heterostructures is ideally compatible with the approach to creating a tunable microcavity structure using an external mirror, providing the most efficient single-photon generation to date.

	\backmatter
	
	\bmhead{Additional information}
	
	Additional data on this work are available with personal requirement.
	
	\bmhead{Acknowledgments}
	
	The authors express their gratitude for partial support of this work within the framework of the State assignment of the Russian Federation No. FFUG-2024-0043.
	
	
	\bmhead{Competing interests}
	The authors declare no competing interests.

	\section*{Methods}
	
	\subsection{Sample designing and fabrication}
	
	The heterostructures for the SPSs were grown by MBE on an epi-ready GaAs(001) substrate. They contain 25 pairs of $\lambda/4$ Al$_{0.9}$Ga$_{0.1}$As/GaAs layers, forming the lower DBR, a linearly graded In$_x$Ga$_{1-x}$As MBL ($x$ varied from 0.05 to 0.43) with a thickness of $\sim 1.1~\mu$m, a layer of self-organized InAs QDs on top of the graded layer and then a 0.2~$\mu$m-thick In$_{0.28}$Ga$_{0.72}$As uniform cap layer. The thickness of both the MBL and the capping layer was chosen to ensure the correct $3\lambda$ cavity length and the location of the QDs at the antinode of the cavity mode in the final structure. Details of MBE growth and characterization of similar heterostructures can be found  in \cite{Sorokin2024, Lakuntsova2026, Sorokin2025, Veretennikov2025}.
	
	QDs emitting in C-band tend to elongate in the $[1\bar{1}0]$ direction due to asymmetric migration of In atoms on the surface, which can lead to the formation of "quantum dashes" \cite{Sorokin2025}. This tendency could be partly suppressed by introducing a thin GaAs sublayer before InAs QDs deposition \cite{Khan2014}. Typically, the average surface density of QDs was in the mid of $10^{10}$ cm$^{–2}$, and their spatial distribution across the substrate was non-uniform, even on a scale of several micrometers. Furthermore, the lateral dimension and height of the QDs also vary, leading to variations in their emission wavelength in the range from 1350 to 1570~nm. High-resolution cross-sectional transmission electron microscopy (TEM) images of individual QDs emitting around 1550~nm show a lateral dimensions $L$ of 30~nm and a heights $H$ of 5-6~nm. The estimated density of such QDs is in the mid of $10^8$~cm$^{–2}$, as reported in \cite{Khan2014, Wyborski2023}.
	
	After growth by MBE, periodically arranged pillars with diameters ranging from 1 to 5 µm were fabricated using photolithography and reactive ion-plasma etching. Then a high-performance thin DBR was fabricated by ion-assisted reactive magnetron sputtering to finish the SPSs. The deposited structure consisted of a 40-nm Si layer followed by two pairs of Si/SiO$_2$. The role of the thin Si layer is to precisely adjust the optical cavity length, which is possible due to the small difference between the Si and In$_{0.28}$Ga$_{0.72}$As refractive indices.
	
The fabrication process was non-deterministic in terms of the mutual spatial and spectral positions of the produced microresonator modes and QDs. For this reason, the parameters of SPSs vary widely, and only a few of the hundreds realize the full potential of the proposed design. Despite this, we have identified several SPSs whose parameters are quite close to theoretical expectations. We consider these structures to be the most representative of our concept. We also note that a minor improvement in parameters and a significant improvement in reproducibility can be achieved using deterministic fabrication processes, such as \textit{in situ} lithography \cite{Somaschi2016}.

	\subsection{Finite-difference time-domain simulations}
	
	The performance of the SPS with two Si/SiO$_2$ pairs and 25 GaAs/AlGaAs pairs in the upper and lower DBRs was evaluated for a numerical aperture of $\text{NA} = 0.68$, which corresponds to our experimental conditions. The simulations were performed using the finite-difference time-domain (FDTD) method for $\lambda/4$ layer thicknesses in both DBRs, with the QD located at the antinode of the optical mode in the $3\lambda$ cavity. In the main text, the parameters of the composite microcavity were evaluated for an idealized design. The range of possible deviations in realistic micropillar structures was determined by taking into account several factors. The micropillar may have some tilt and sidewall roughness; deviations in the DBR layer thicknesses from $\lambda/4$ may also occur. The scale of such errors was estimated based on the analysis of reflectance spectra and SEM images of test samples obtained using the same fabrication procedure. By modeling a non-ideal structure with typical deviations, we found that their impact on the quality factor and Purcell factor is relatively small, reducing them to $Q \approx 1100$ and $F_p \approx 2.8$, respectively. At the same time, the extraction efficiency at a given numerical aperture was more sensitive, decreasing to $\eta_\text{out}\approx45\%$. All these values still correspond to the optimal location of the QD at the center of the micropillar, which is ensured in the experiment by the selection of devices with high single-photon emission brightness with location of the emission spot in the center of a pillar apex. The best devices are expected to achieve $\eta_\text{out}$ close to 65\% predicted for the idealized geometry.

	\subsection{Optical measurements}
	
\begin{figure}[t]
		\centering
		\includegraphics[width=0.7\textwidth]{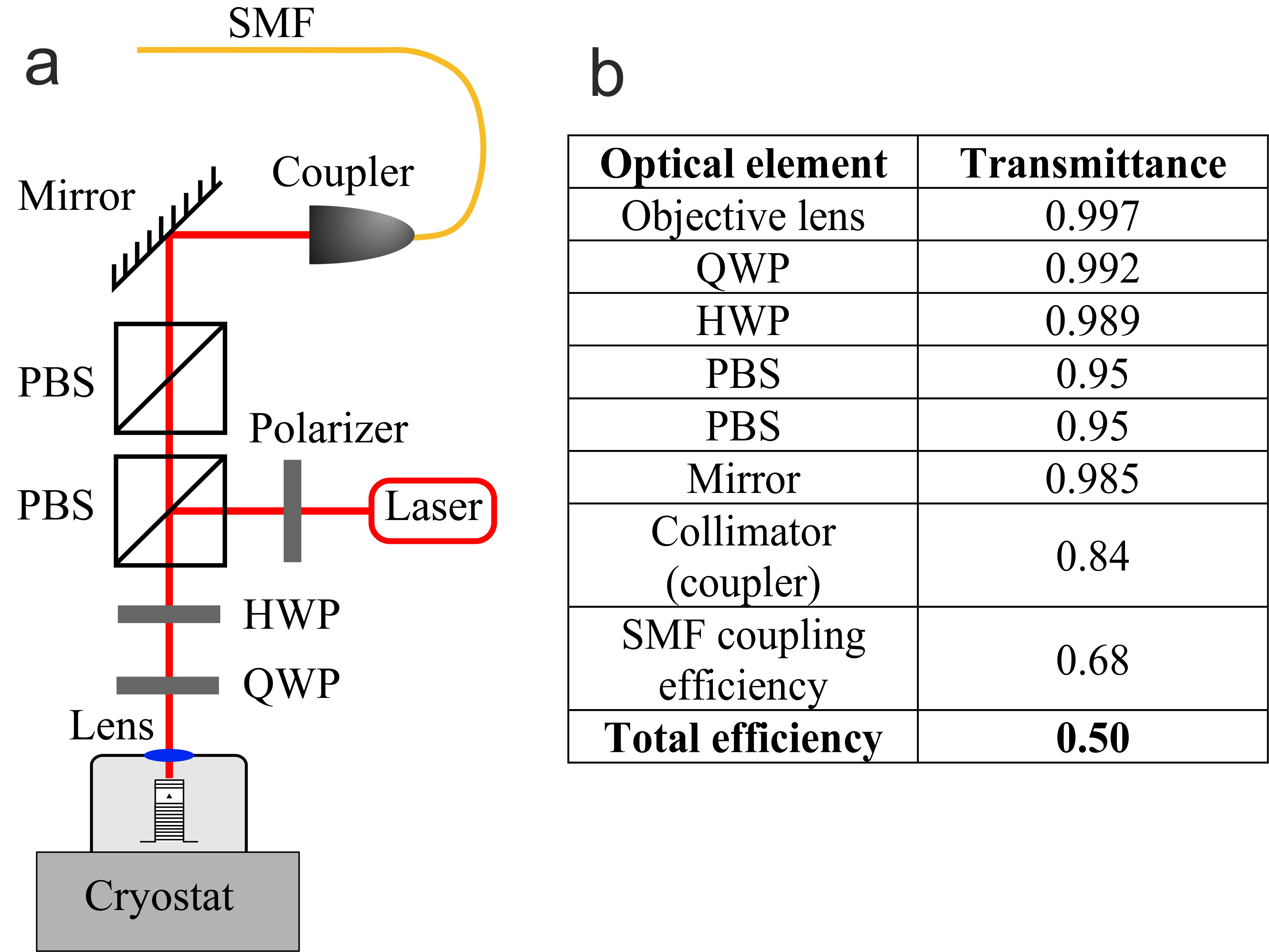}
		\caption{ $\vert$ \textbf{Methods. Details of optical measurements.}
			\textbf{a}, Schematic of the optical setup for SPS cross-polarized resonant excitation. \textbf{b}, Light transmittance through the optical elements used. }
		\label{figM1} 
	\end{figure}

	The optical setup for resonant excitation of a QD is shown in Fig.~\ref{figM1}a. The sample is mounted in a continuous-flow helium cryostat maintaining a temperature in the range of 8–300~K. The laser excitation  pulses (repetition rate 78.3~MHz) are generated by a femtosecond optical parametric oscillator system. A $4f$ pulse shaper \cite{Monmayrant2010} enables precise wavelength tuning and spectral narrowing to match individual QDs for efficient resonant pumping. The laser pulses pass through a linear polarizer and a polarizing beam splitter (PBS), then are focused onto the apex of a selected micropillar resonator using a high-NA aspherical objective lens ($\text{NA} = 0.68$) with an anti-reflective coating, integrated into the cryostat. Single-photon emission from the composite micropillar is collected by the same lens and directed through two cascaded polarizing beamsplitters, which suppress the orthogonally polarized laser beam with an extinction ratio of $\approx 10^5$. A half-wave plate (HWP) and a quarter-wave plate (QWP) are used sequentially to control the orientation of the polarization vector in the linearly polarized excitation light relative to the microcavity crystalline axes, while compensating for the parasitic polarization ellipticity introduced by the optical components. Single-photon emission, isolated via cross-polarization filtering, is coupled into a single mode fiber (SMF) using an adjustable-focus collimator (coupler). The main single photon emission parameters obtained from the SPS \textbf{A} measurements are presented in Table~\ref{tab:parameters_list}; details on the measurements and analysis are given below.
	
	\begin{table}[h]
		\caption{Measured parameters of the SPS \textbf{A} with a composite semiconductor-dielectric micropillar resonator. All parameters are provided at the optimal operational temperature under resonant $\pi$-pulse excitation, unless stated otherwise.}
		\label{tab:parameters_list}
		\begin{tabular}{|l|c|}
			\hline
			Trion line wavelength & 1551.1 nm \\
			\hline
			Optimal operating temperature & 16 K \\
			\hline
			Trion state lifetime & 410 ps \\
			\hline
			End-to-end efficiency \(\eta_{\text{end}}\) & (11.0 ± 0.3)\% \\
			\hline
			Polarized first lens brightness \(\eta_{\text{FL\,pol}}\) & (22.0 ± 1.5)\% \\
			\hline
			Unpolarized first lens brightness \(\eta_{\text{FL\,unpol}}\) & (44 ± 3)\% \\
			\hline
			Stability of QD bright state \(\eta_{\text{blink}}\) & (95.0 ± 0.9)\% \\
			\hline
			Excitation efficiency \(\eta_{\text{exc}}\) & (77 ± 4)\% \\
			\hline
			Light extraction efficiency \(\eta_{\text{out}}\) & (60 ± 8)\% \\
			\hline
			\(g^{(2)}(0)\) at the \(\pi\)-pulse excitation & 0.043 ± 0.002 \\
			\hline
			\(g^{(2)}(0)\) with the 30\% signal loss\footnotemark[1] & 0.030 ± 0.002 \\
			\hline
			HOM interference visibility \(V_{\text{HOM}}\) at 16 K & 22\% \\
			\hline
			HOM interference visibility \(V_{\text{HOM}}\) at 8.3 K & 38\% \\
			\hline
		\end{tabular}
		\footnotetext[1]{This signal loss corresponds to the reduced excitation power.}
	\end{table}
	
	\subsubsection{Brightness and signal loss analysis}
	
	The optical losses introduced by each component used were characterized using laser light at the QD emission wavelength. The SMF coupling efficiency was determined from the ratio of the laser signals collected through the SMF to the signals collected through the multimode fiber capturing the total radiation after the collimator. For accurate calibration, we used laser light, coupled and emitted from the micropillar. It was filtered from directly reflected laser light by polarization, since laser photons interacting with the microresonator exhibit a slight polarization rotation \cite{Ollivier2020}. The optical properties of this light, closely resembling single-photon emission, allow for reliable calibration of the coupling efficiency \cite{Hilaire2018}. The obtained transmittance values for all components and the coupling efficiency are presented in Fig.~\ref{figM1}b. They correspond to the overall efficiency of the optical setup from the signal incident on the first lens to the signal in the fiber of $\eta_\text{opt} = (50\pm2)\%$.
	
	For brightness measurements, a single-mode optical fiber was directly connected to a SNSPD with a detection efficiency of up to $\eta_\text{det} = (92\pm3)\%$ (see below) and a dark count rate below 300 Hz. The electrical signal was then analysed using a Time Tagger. Rabi oscillations in the PL emission were measured with pulse areas increasing up to $6\pi$. They were modeled numerically by solving the optical Bloch equations, assuming a power-law dependence of the pure dephasing rate and taking into account possible multiphoton generation for even-$\pi$ pulses \cite{Krainov2025}.
	
	\subsubsection{Single-photon purity}
	
	The single-photon purity was assessed by measuring the auto-correlation function $g^{(2)}(t)$ in a HBT setup with a fiber beam splitter using two SNSPDs. The obtained histogram $c_\text{HBT}(t)$ was modeled under the assumption that the detected signal contains strong single photon emission of blinking single QD with a lifetime $\tau$ and some parasitic weak emission source with another lifetime $\tau_2$, which is necessary to describe the different width of the zero peak compared to the others. The auto-correlation of the parasitic minor component was neglected as a quantity of second-order smallness. That leads to the following fitting function:
	
	\begin{align}
		\label{eq:g2_fit}
		c_\text{HBT}(t)=\left[c_0 + 
		\frac{A}{2\tau}\cdot f_\text{blink} (t) \cdot \right.
		&\sum_{i \ne 0} e^{-\frac{|t-iT|}{\tau}} + \nonumber \\ 
		\frac{B}{2 (\tau + \tau_2)} \cdot
		&\left.\sum_i \left(e^{-\frac{|t-iT|}{\tau}} + e^{-\frac{|t-iT|}{\tau_2}} \right)
		\right] \ast I_\text{instr}(t),
	\end{align}
	where $c_0$ is the background level, $i$ is the peak number, $T$ is the laser pulse repetition period, $A$ and $B$ are normalization constants for emission components and $f_\text{blink}(t)$ describes the blinking-induced bunching of the $g^{(2)}(t)$ function:
	\begin{equation}
		f_\text{blink}(t) = 1 + \left(\frac1{\eta_\text{blink}} - 1 \right)\cdot 
		e^{-\frac{|t|}{\tau_\text{blink}}}.
		\label{eq:g2_blinking}
	\end{equation}
	Here $\tau_\text{blink}$ is the typical bunching time, and $\eta_\text{blink}$ gives the QD efficiency, i.e. the probability to find the QD in a bright state at the arbitrary moment. Eq.~\ref{eq:g2_fit} contains also the convolution with the Gaussian instrumental function $I_\text{instr}(t)$. The normalization constants obtained in the fitting procedure give the side peak areas $(A+B)$ and zero peak area $B$, used to evaluate $g^{(2)}(0)=B/(A+B)$. 
	
	\subsubsection{Indistinguishability}
	
	Photon indistinguishability was characterized by measuring the in-fiber Hong-Ou-Mandel (HOM) interference visibility $V_\text{HOM}$ for different inter-photon delays and temperatures. QD emission, after cross-polarization filtering with a polarizing beam splitter, was split into two channels and then coupled to fibers of different lengths to achieve the desired delay $T_\text{HOM}$. The signals were then combined in a fiber optic coupler in either the same or orthogonal polarizations, and the delay statistics $c_\text{HOM}(t)$ were collected with two SNSPDs.
	
	During the HOM measurements, each photon could pass through the long or short interferometer arm with the probabilities determined by the first beam splitter's transmission and reflection coefficients, $T_\text{BS\,1}$ and $R_\text{BS\,1}$. If no interference occurs, i.e. for orthogonal polarization in interferometer arms, each photon will then pass to the detectors with the probabilities determined by the second beam splitter's coefficients, $T_\text{BS\,2}$ and $R_\text{BS\,2}$. In this case the fitting function $c_\text{HOM}(t)$ could be directly expressed in terms of the $g^{(2)}(0)$ function:
	\begin{align}
		c_{\text{HOM}\,\perp}(t) = & c_0 + C \cdot \left[
		\left( T_\text{BS\,1}^2 + R_\text{BS\,1}^2 \right) T_\text{BS\,2} R_\text{BS\,2}\cdot g^{(2)}(t) + \right. \nonumber \\ & \left.
		T_\text{BS\,1} R_\text{BS\,1} R_\text{BS\,2}^2 \cdot g^{(2)}(t+T_\text{HOM}) + 
		T_\text{BS\,1} R_\text{BS\,1} T_\text{BS\,2}^2 \cdot g^{(2)}(t-T_\text{HOM}) \right],
		\label{eq:HOM_cross_fit}
	\end{align}
	where $c_0$ is the background level and $C$ is a normalization constant. This equation describes the photon counting statistics in an HOM measurement for an arbitrary delay $T_\text{HOM}$, but in the case of small delays, $T_\text{HOM} < T$, the excitation pulse duplication is necessary and should be taken into account in the $g^{(2)}(t)$ function. That could be done either by simulation or by direct measurement of the $g^{(2)}(t)$ function in the HOM setup with one interferometer arm shut down.
	
	For the co-polarized measurement of photons with a time-dependent wave packet overlap $M(t)$ using an imperfect interferometer with classical interference visibility $(1-\epsilon)$, the detection probability of a photon pair arriving at the interferometer will be proportional to the expression \cite{Santori2002}:
	\begin{equation}
		p(t)=T_\text{BS\,2}^2 + R_\text{BS\,2}^2 - 2 \cdot (1-\epsilon)^2 \cdot M(t) \cdot T_\text{BS\,2} \cdot R_\text{BS\,2},
	\end{equation}
	where the last term represents the contribution from single-photon interference. Assuming that the wavepacket overlap decays exponentially as $M(t) = M_0 \cdot \exp\left(-\frac{2\cdot |t|}{T^*_2}\right)$ with a pure dephasing time $T^*_2$ \cite{Bylander2003}, we obtain the fitting function for the co-polarized HOM measurement:
	\begin{equation}
		c_{\text{HOM}\,\parallel}(t) = c_{\text{HOM}\,\perp}(t) - \left[ 
		2 (1-\epsilon)^2 M_0 \cdot
		T_\text{BS\,1} R_\text{BS\,1} T_\text{BS\,2} R_\text{BS\,2} \cdot 
		\frac{C}{2\tau} \cdot	
		e^{-\frac{2\cdot |t|}{T_2}}
		\right] \ast I_\text{instr}(t).
		\label{eq:HOM_co_fit}
	\end{equation}
	Here the coherence time $T_2$ is defined as $\frac1{T_2} = \frac1{T^*_2} + \frac1{2 T_1}$, and all other parameters are identical to those in Eqs.~\ref{eq:g2_fit} and \ref{eq:HOM_cross_fit}.

	\subsection{SNSPD efficiency}
	
	The photon detection rate was measured for the SNSPD operating in two modes: with current values of $I_1 = 18.5$~$\mu$A and $I_2 = 18.0$~$\mu$A. The first mode provides higher detection efficiency of $\eta_\text{det} = (92\pm3)\%$, but increases sensitivity to electromagnetic noise from the power supply circuits. The second mode ensures a more stable system operation at the expense of lower $\eta_\text{det} = (88\pm2)\%$. The values of $\eta_\text{det}$ were measured as the probability of detecting an attenuated laser pulse, defined as the ratio of the detection count rate to the incidence count rate. For this measurement, the resonant laser was sent to the SNSPD through an attenuator to achieve a detection count rate similar to that observed during the SPS characterization. This was necessary because the detection efficiency generally depends on the wavelength and signal level. The incidence count rate was then evaluated from the known mean power and pulse repetition period of the laser, taking into account the Poissonian distribution of the photon number in the laser pulses.
	
	The values of the detection count rates in the main text are given for the first mode, since the higher $\eta_\text{det}$ makes the results more representative of the SPS properties. We note that, within the error margins, the results for both measurement modes are in good agreement; therefore, we used the averaged results to estimate the first-lens brightness and end-to-end efficiency values provided in the main text and in Table~\ref{tab:parameters_list}.

	\bibliography{sn-bibliography}
	
\end{document}